\documentclass[twocolumn]{aastex701}

\usepackage{color}
\usepackage{amsmath,amssymb}

\DeclareMathOperator{\sech}{sech}
\shorttitle{Reconnection vs Turbulence in Blazars}
\shortauthors{Zhang et al.}

\begin{document}

\title{Energy-Dependent Polarization Angle Variability as a Robust Diagnostic for Blazar Flaring Mechanisms}

\correspondingauthor{Haocheng Zhang}

\author[0000-0001-9826-1759]{Haocheng Zhang}
\affiliation{University of Maryland Baltimore County\\
Baltimore, MD 21250, USA}
\affiliation{NASA Goddard Space Flight Center\\
Greenbelt, MD 20771, USA}
\email{haocheng.zhang@nasa.gov}

\author[0009-0008-0415-7263]{Benjamin de Jonge}
\affiliation{Department of Physics, Washington University in St Louis\\
St. Louis, MO 63130, USA}
\email{bdejonge@wustl.edu}

\author[0000-0002-1853-863X]{Manel Errando}
\affiliation{Department of Physics, Washington University in St Louis\\
St. Louis, MO 63130, USA}
\email{errando@wustl.edu}

\author[0000-0001-5278-8029]{Xiaocan Li}
\affiliation{Theoretical Division, Los Alamos National Lab \\
Los Alamos, NM 87545, USA}
\email{xiaocanli@lanl.gov}

\author[0000-0003-4315-3755]{Fan Guo}
\affiliation{Theoretical Division, Los Alamos National Lab \\
Los Alamos, NM 87545, USA}
\affiliation{New Mexico Consortium, Los Alamos, NM 87544, USA}
\email{guofan@lanl.gov}



\begin{abstract}
Identifying the physical mechanism driving blazar flares remains a central challenge in high-energy astrophysics. We show that the energy dependence of the standard deviation of the polarization angle variability ($\sigma_\text{PA}$) provides a powerful and robust discriminator of blazar flaring mechanisms. Using particle-in-cell-integrated polarized radiative transfer simulations, we perform to-date the most rigorous statistical analyses of polarization variability. We demonstrate that magnetic reconnection and magnetized turbulence imprint qualitatively distinct energy dependence of $\sigma_\text{PA}$ that directly reflect their different magnetic field evolution and particle transport. Reconnection predicts higher $\sigma_\text{PA}$ with higher photon energy till the synchrotron spectral peak, whereas turbulence produces nearly flat $\sigma_\text{PA}$ across the synchrotron spectral component. These trends are resilient to realistic observational limitations. Applying our results to optical and IXPE data of Mrk~421 and 1ES~1959+650, we find strong evidence for reconnection-driven flares embedded in a turbulent blazar zone. Energy-dependent $\sigma_\text{PA}$ emerges as a decisive new probe of particle acceleration in relativistic jets.
\end{abstract}



\section{Introduction} \label{sec:intro}

Blazars are known to be highly variable across the entire electromagnetic spectrum. The fastest flares can be as brief as a few minutes \citep{Ackermann2016,Aharonian2007,Albert2007}. The strong variability indicates extreme particle acceleration in very localized regions, often referred to as the blazar zone. Blazar flares have been considered as potential sources of extragalactic cosmic rays and neutrinos \citep{IceCube2018,Murase2018}. Despite extensive observational and theoretical efforts, the dominant particle acceleration mechanism responsible for blazar flares remains unsettled.

Shock acceleration, magnetic reconnection, and turbulence are frequently invoked to explain particle acceleration and nonthermal radiation in Blazars. Extensive theoretical and numerical efforts have shown that each of them can produce a nonthermal electron population with power-law energy spectra and variable multi-wavelength emission broadly consistent with observations \citep{Marscher1985,Drury1999,Spitkovsky2008,Drury2012,Guo2014,Guo2024,Sironi2025,Zhdankin2017,Comisso2018,Li2023,French2023}. Crucially, however, these mechanisms operate under fundamentally different plasma conditions and give rise to qualitatively distinct jet dynamics and magnetic field evolution. Identifying the dominant mechanism therefore provides direct insight into jet composition, energy dissipation, and the ability of relativistic jets to accelerate ultra-high-energy particles. 

Multi-wavelength polarimetry has emerged as a powerful diagnostic of magnetic field structure in blazar jets \citep{Boettcher2019,ZHC2019c}. Recent IXPE observations of high-synchrotron-peaked (HSP) blazars have revealed systematically higher polarization degrees in X-rays than in the optical or radio bands \citep{Liodakis2022,DiGesu2022}. This has been interpreted under a simplified shock acceleration scenario: high-energy electrons radiating in X-rays are confined near the shock front due to fast cooling, where the magnetic field is more ordered, whereas lower-energy electrons are advected downstream into a more turbulent field, reducing the optical and radio polarization \citep{Angelakis2016,Liodakis2022}. However, similar energy-stratified polarization degrees can arise in alternative scenarios. For example, diffusion models allow low-energy electrons to populate a larger, more turbulent region, while high-energy electrons again remain spatially confined due to rapid cooling, producing higher polarization at higher photon energies \citep{ZHC2024}. Consequently, polarization degree alone does not uniquely identify the flaring mechanism. A more discriminating observable is required.

Polarization variability, particularly the standard deviation of the variable polarization angles during blazar flares, $\sigma_\text{PA}$\footnote{Due to the $180^{\circ}$ ambiguity of the polarization angle, $\sigma_\text{PA}$ is actually the circular standard deviation.}, offers a more powerful diagnostic. Unlike polarization degree, $\sigma_\text{PA}$ directly reflects how much the polarization angle varies during blazar flares, tracing magnetic field evolution and the tightly coupled energy dissipation and particle transport. Its energy dependence encodes how electrons of different energies evolve and sample magnetic field structures across the flaring region. Observations have reported polarization variations in radio, infrared, optical, and recently X-rays \citep{Angelakis2016,Marscher2010,Jorstad2017,DiGesu2023}. Yet, interpreting these measurements requires self-consistent physical models that connect plasma dynamics, particle acceleration, and radiation transfer. In particular, a predictive understanding of how $\sigma_\text{PA}$ depends on photon energy is still lacking.

Particle-in-cell (PIC) simulations provide a first-principles framework to address this problem. By self-consistently evolving particles and magnetic fields, PIC simulations naturally capture particle acceleration and plasma dynamics in magnetic reconnection, and turbulence \citep{Sironi2009,Sironi2014,Guo2014,Guo2016,Guo2021,Zhdankin2017,Comisso2019}. Pioneering works have already hinted some distinctive observable signatures, such as fast flares and polarization angle swings \citep{ZHC2018,ZHC2022,Hosking2020}. But a comprehensive comparison of polarization variability statistics across different flaring mechanisms has been missing. In particular, the diagnostic power of $\sigma_\text{PA}$ has not been fully exploited.

In this paper, we perform unprecedentedly detailed analyses to explore energy dependence of polarization variability arising from magnetic reconnection and turbulence, focusing on $\sigma_\text{PA}$ as a robust discriminator of blazar flaring mechanisms. Using PIC-integrated polarized radiative transfer simulations, we show that magnetic reconnection and turbulence imprint qualitatively distinct and observationally robust trends in $\sigma_\text{PA}$ across photon energies. Reconnection predicts a sharp rise in $\sigma_\text{PA}$ till the synchrotron peak, whereas turbulence produces a nearly energy-independent $\sigma_\text{PA}$. These signatures persist under realistic observational conditions, including limited cadence, sensitivity, and asynchronous multi-band coverage. Applying our results to recent optical and IXPE polarization measurements of Mrk~421 and 1ES~1959+650, we demonstrate that $\sigma_\text{PA}$ provides a decisive and physically motivated diagnostic for identifying the dominant blazar flaring mechanism. Section~\ref{sec:results} presents our simulation setups and results, Section~\ref{sec:interpretation} explains the physical causes of energy dependence of polarization variability, Section~\ref{sec:application} describes the application to optical and IXPE observations, and Section~\ref{sec:summary} summarizes and discusses our findings.

\section{Results \label{sec:results}}

\begin{figure}
\centering
\includegraphics[width=0.99\linewidth]{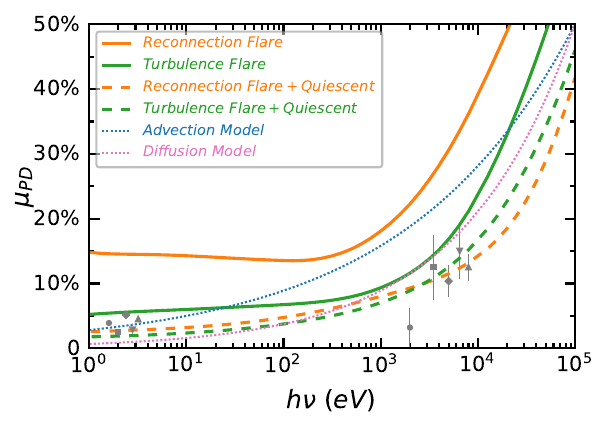}
\includegraphics[width=0.99\linewidth]{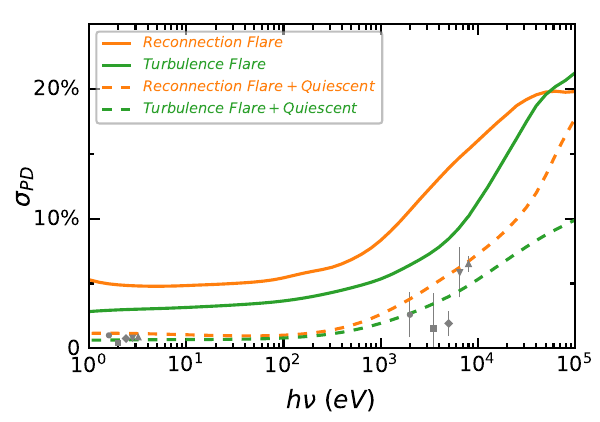}
\includegraphics[width=0.99\linewidth]{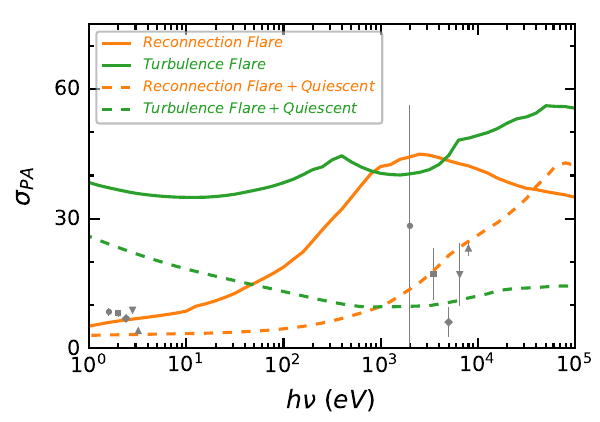}
\caption{Energy dependence of the average ($\mu_\text{PD}$, upper panel) and standard deviation ($\sigma_\text{PD}$, middle panel) of polarization degree, and standard deviation ($\sigma_\text{PA}$, lower panel) of the polarization angle. Orange curves represent reconnection and green curves are turbulence. Gray data points are optical and X-ray polarization of Mrk~421 and 1ES~1959+650. All data points are in the same optical or IXPE band, but they are deliberately spread over the optical and X-ray bands for clarity.}
\label{fig:polarization}
\end{figure}

We use PIC simulation setups of magnetic reconnection and turbulence that are similar to our previous works \citep{ZHC2020,ZHC2023}, but generalized to model HSPs. Details of the setup and parameter survey are in Appendix~\ref{sec:simulation}. Each simulation represents a single, localized flaring region that dominates the blazar zone emission during its flaring phase, analogous to a one-zone model but based on first-principles plasma simulations. In reality, blazars can have multiple flaring regions \citep{Chen2014,Marscher2014,deJonge2026}; moreover, telescopes have limited sensitivity, with fluxes that fall below a certain sensitivity limit being typically undetectable. To incorporate these effects, we normalize the maximum flux in the IXPE band for all simulation runs to the same value of $F_{peak}=6F_0$, in arbitrary flux units, and select a flux thresholds of $F_0$ or $2F_0$ in the same units to generate our simulated data sets (referred to as threshold 1 or 2 in the following). We only consider emission above the selected threshold, thus our pseudo-observation of HSPs corresponds to the high-flux state in X-rays.

Telescopes often have constrained observational cadences due to operational conditions (weather, operational modes, etc.). In addition, their limited sensitivity forces observers to accumulate data over times that may be longer than the dynamical timescales of the system in order to collect enough photon statistics and sufficient signal-to-noise ratio to resolve the polarization properties of the source. A changing polarization angle, if unresolved in time, will reduce the measured polarization \citep{ZHC2021,Errando2024,DiGesu2023}. In order to mimic the limited sensitivity, cadence and time resolution of real observational data sets, we generate two different sets of mock observations (or pseudo-observations) from the simulated light curves by averaging the Stokes parameters over 5 and 30 simulation steps from PIC-integrated radiation transfer simulations\footnote{One simulation step is $\sim 0.005\tau_{lc}$, where $\tau_{lc}$ is the light crossing time of the simulation box. See Appendix~\ref{sec:simulation}.} (referred to as cadence 5 or 30 in the following), which naturally include the effects of depolarization due to limited time resolution. We require those 5 or 30 snapshots to be continuously higher than the threshold selected above for the resulting observed Stokes parameters to be treated as one observational data point, mimicking the limited sensitivity and cadence under various realistic observational conditions. Details of our pseudo-observation method are given in Appendix~\ref{sec:analysis}. As shown in Appendix~\ref{sec:robustness}, these arbitrary choices of thresholds and cadences do not affect our key results.

We collect all data points in each run (flaring episode) selected by the same or a mixture of thresholds and cadences. Then we calculate the mean and standard deviation of the polarization degree ($\mu_\text{PD}$ and $\sigma_\text{PD}$) and the standard deviation of the polarization angle ($\sigma_\text{PA}$) in all wavelengths in the synchrotron component. These quantities are finally averaged over all runs of reconnection or turbulence, allowing us to isolate systematic energy-dependent behaviors. Our simulations use identical plasma parameters wherever applicable to enable direct comparison between reconnection and turbulence. The resulting polarization variability statistics can be directly confronted with multi-wavelength polarimetric campaigns on one or a number of blazar flaring episodes lasting a few days to weeks. In particular, $\sigma_\text{PA}$ emerges as the most discriminating observable.

Figure~\ref{fig:polarization} shows the energy dependence of the polarization variability. Solid curves represent the single flaring region represented by our simulations, while dashed curves, aiming to fit observational data, include an additional quiescent contribution from the rest of the blazar zone (dashed curves and the fitting are described in more detail in Section~\ref{sec:application}). These curves are derived from the simultaneous data points with threshold 1 and cadence 5 that have the best statistics (see Appendix~\ref{sec:analysis} for details of threshold and cadence choices). But as demonstrated in Appendix~\ref{sec:robustness}, these trends are robust to asynchronous observational data with various sensitivity and cadences.

Both the mean $\mu_\text{PD}$ and standard deviation $\sigma_\text{PD}$ of the polarization degree are nearly flat or rising slowly as a function of observed energy for reconnection and turbulence below the synchrotron spectral peak ($h\nu_\text{peak}$ is at a few hundred eV, see Figure~\ref{fig:xraylcsed}). Starting near the synchrotron peak, both $\mu_\text{PD}$ and $\sigma_\text{PD}$ rise sharply. Overall, reconnection shows higher $\mu_\text{PD}$ and $\sigma_\text{PD}$ than turbulence. However, polarization degree alone remains an ambiguous diagnostic, as its absolute value can be strongly diluted by the quiescent emission from other parts of the blazar zone. The energy dependence of $\mu_\text{PD}$ and $\sigma_\text{PD}$ shows little difference between reconnection and turbulence. Notably, semi-analytical shock and diffusion models (blue \citep{Angelakis2016} and pink \citep{ZHC2024} curves in Figure~\ref{fig:polarization}, respectively) predict a smooth, monotonic increase of $\mu_\text{PD}$ with energy that lacks the sharp transition near the synchrotron peak seen in our simulations. This inconsistency suggests that in our PIC-integrated simulations, radiative cooling alone cannot fully explain the energy-stratified polarization degree.

The most striking and robust difference between flaring mechanisms appears in the energy dependence of $\sigma_\text{PA}$. In turbulence, $\sigma_\text{PA}$ is large at all photon energies, reflecting stochastic magnetic field orientations across a wide range of spatial scales. In contrast, magnetic reconnection predicts a low $\sigma_\text{PA}$ at energies well below the synchrotron peak, followed by a rapid increase toward higher energies. Above the synchrotron peak, $\sigma_\text{PA}$ saturates at values comparable to those produced by turbulence. This behavior directly reflects the transition in particle spatial distributions, from low-energy electrons that distribute nearly homogeneous in plasmoids to high-energy electrons that are confined in rapidly evolving, localized magnetic structures. As a result, $\sigma_\text{PA}$ provides a clear and physically motivated discriminator between reconnection and turbulence.

\section{Physical Interpretation \label{sec:interpretation}}

\begin{figure*}
\centering
\includegraphics[width=0.99\linewidth]{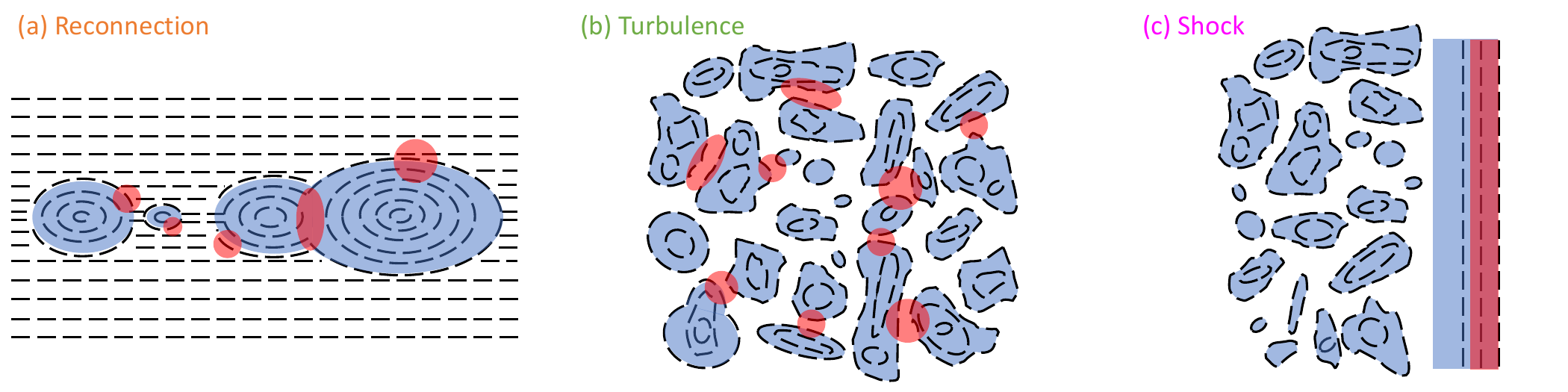}
\caption{A toy model sketch that shows the differences between reconnection, turbulence, and simplified shock scenarios. Dash lines represent the magnetic field lines, blue color represents the spatial distributions of low-energy electrons, and red color represents the spatial distribution of high-energy electrons. In both reconnection and turbulence, high-energy electrons tend to occupy a part of the edge or the merging sites of plasmoids/plasma structures, while low-energy electrons tend to fill up those structures. Under the simplified shock scenario, high-energy electrons occupy the shock front, while low-energy electrons can exist in both the shock front and turbulent downstream. Note that more physical shock simulations suggest an overall turbulent environment, similar to the turbulence scenario, rather than the simplified shock scenario here.}
\label{fig:sketch}
\end{figure*}

\begin{figure*}
\centering
\includegraphics[width=0.99\linewidth]{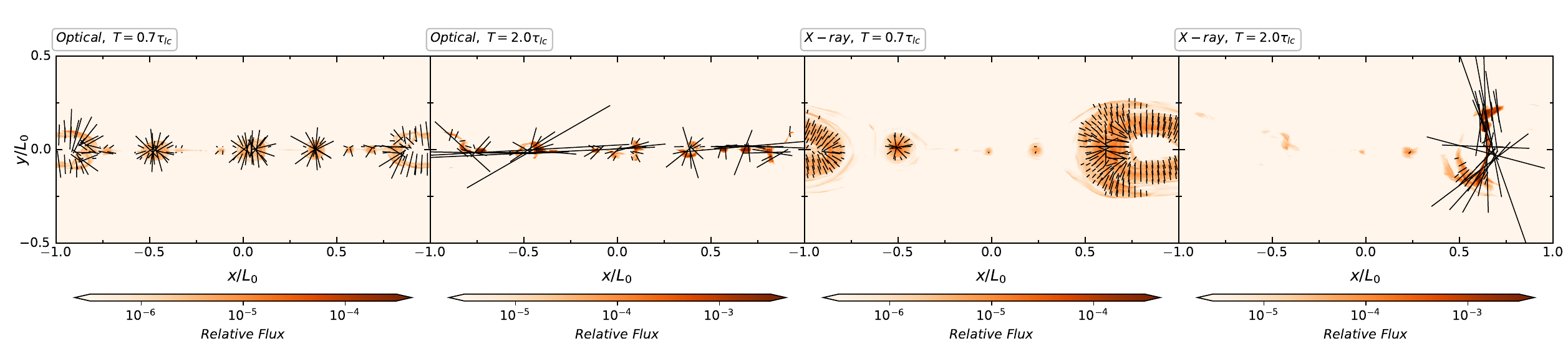}
\includegraphics[width=0.99\linewidth]{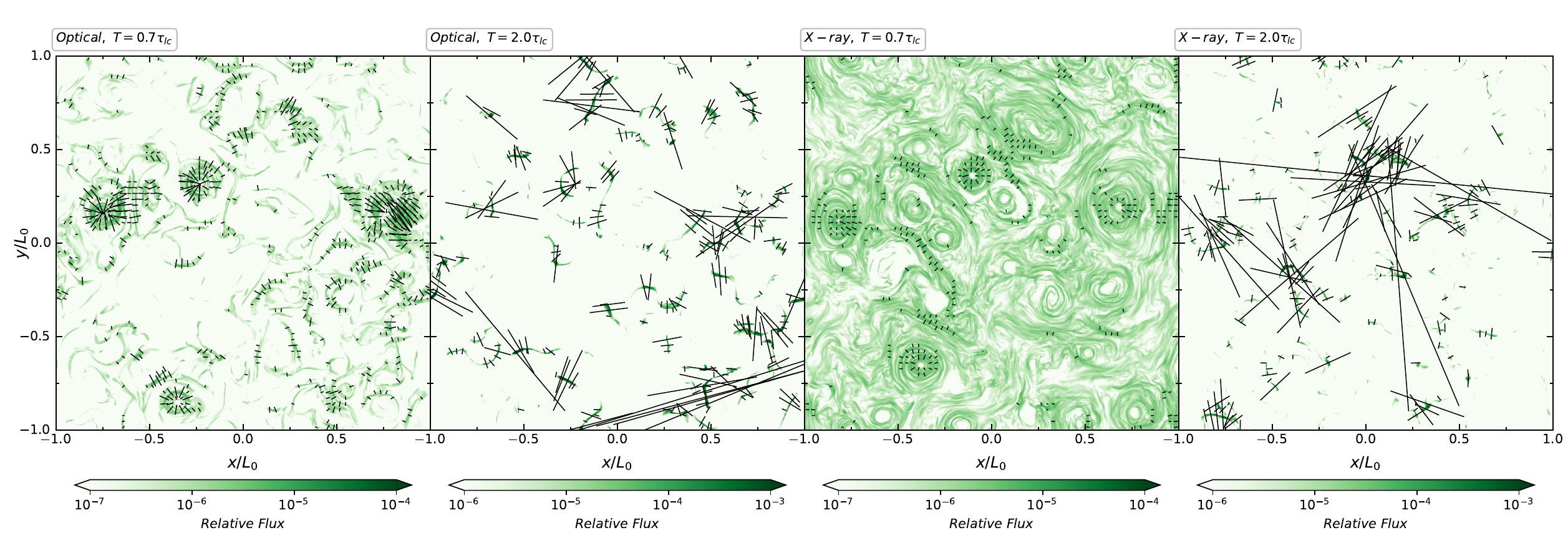}
\caption{Polarized emission maps of reconnection (upper panels) and turbulence (lower panels). The first and third panels in each row are the optical emission map in an early and a late snapshot, respectively, while the second and fourth panels are the IXPE band. The orange and green colors represent the total luminosity distribution of reconnection and turbulence, respectively. The length and direction of black lines represent the local polarization degree and angle, respectively.}
\label{fig:emissionmap}
\end{figure*}

The distinct energy-dependent polarization variability described in Section~\ref{sec:results} is rooted in the co-evolution of nonthermal particles and magnetic field in reconnection and turbulence scenarios. Figures~\ref{fig:sketch} and \ref{fig:emissionmap} provide schematic and simulation-based illustrations of these differences. In both reconnection and turbulence, synchrotron emission is dominated by localized plasma structures, such as plasmoids, plasmoid mergers, or turbulent eddies, rather than a homogeneous flaring region. These structures act as discrete ``emission patches'', each characterized by its own magnetic field orientation and particle distribution. The observed polarization results from the superposition of radiation from many such patches. How electrons of different energies populate these patches therefore determines the energy dependence of polarization variability.

In magnetic reconnection, electrons before the cooling break (blue color in Figure~\ref{fig:sketch}a), which corresponds to the synchrotron spectral peak, do not suffer from significant cooling, and they will occupy nearly the entire volume of plasmoids (Figure~\ref{fig:emissionmap} first and third panels in the upper row). These quasi-coherent structures have similar and stable field orientations, exhibiting a low and stable polarization degree and a nearly steady polarization angle perpendicular to the current sheet in reconnection \citep{ZHC2024b}. As a result, both $\mu_\text{PD}$ and $\sigma_\text{PD}$ remain low, and $\sigma_\text{PA}$ is strongly suppressed at low photon energies. With increasing electron energy (but still below the cooling break), particle distributions become progressively more inhomogeneous within plasmoids, introducing slightly stronger polarization angle variability. This leads to a gradual rise of $\sigma_\text{PA}$ as the photon energy approaches the synchrotron peak. We note that although our PIC simulations are 2D, 3D reconnection simulations also show flux ropes \citep{Li2019,Guo2015,Guo2021,Zhang2021,Bacchini2025,French2026}, similar to plasmoids in 2D, which should result in similar polarization signatures.

Near and above the synchrotron peak, reconnection-driven polarization behavior changes qualitatively. High-energy electrons (red color in Figure~\ref{fig:sketch}a) suffer from strong radiative cooling and remain confined to localized regions near plasmoid edges or merger sites, where the magnetic field is more ordered but varies rapidly in time and space (Figure~\ref{fig:emissionmap} second and fourth panels in the upper row). Emission from these regions exhibits higher polarization degrees but samples different magnetic field orientations across distinct plasmoids and merger events. When averaged over multiple emission patches, this leads to enhanced polarization degree variability and a sharp increase in $\sigma_\text{PA}$. At even higher energies, $\sigma_\text{PA}$ saturates, reflecting the fact that the emitting electrons are always confined to localized and dynamically evolving regions. The sharp rise in the energy dependence of $\sigma_\text{PA}$ is thus the defining signature of reconnection scenarios.

In contrast, turbulence-driven flares exhibit fundamentally different $\sigma_\text{PA}$. Magnetized turbulence generates irregular magnetic field structures across a wide range of spatial scales, without any preferred orientations. While high-energy electrons (red color in Figure~\ref{fig:sketch}b) are still spatially confined in active particle acceleration sites due to cooling, resulting in higher $\mu_\text{PD}$ and $\sigma_\text{PD}$ than low-energy electrons (blue color in Figure~\ref{fig:sketch}b), magnetic fields sampled by electrons of any energy do not have preferred orientations. Consequently, emission patches contributing at both low and high photon energies exhibit a broad distribution of polarization angles. leading to persistently large $\sigma_\text{PA}$ across the entire spectrum, with only weak  energy dependence.

\begin{figure}
\centering
\includegraphics[width=0.99\linewidth]{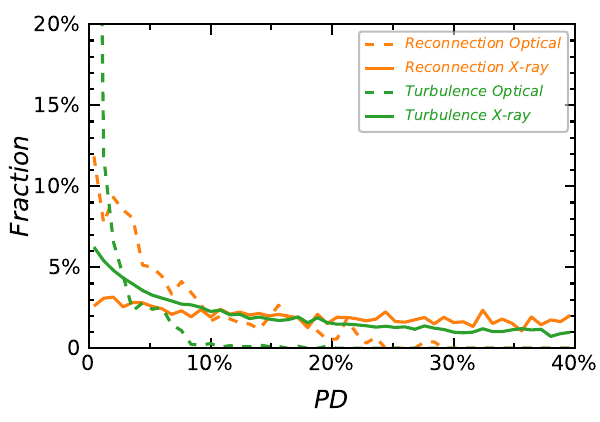}
\includegraphics[width=0.99\linewidth]{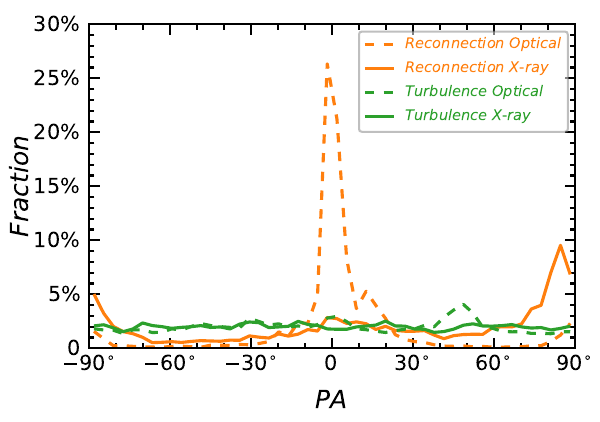}
\caption{Histograms of contribution to total luminosity by plasma structures of various polarization degree (upper panel) and angle (lower panel) in reconnection (orange) and turbulence (green). Dashed lines represent optical emission and solid lines are for the IXPE band.}
\label{fig:statistics}
\end{figure}

These differences are further quantified by examining the statistical distributions of polarization properties from individual emission patches (Figure~\ref{fig:statistics}). In both scenarios, patches emitting above the synchrotron peak reach higher polarization degrees, driving the rise in $\mu_\text{PD}$ and $\sigma_\text{PD}$. However, only reconnection shows a strong concentration of polarization angles around a preferred orientation for patches emitting below the synchrotron peak. This concentration leads directly to low $\sigma_\text{PA}$ at low energies. In turbulence, the polarization angle distribution remains broad at all energies, explaining the weak energy dependence of $\sigma_\text{PA}$. The contrast highlights why $\sigma_\text{PA}$ is uniquely sensitive to the co-evolution of particles and magnetic fields.

Lastly, we discuss the expected polarization variability for the simplified shock scenario that is often used to explain higher polarization degree in IXPE band than that in the optical band \citep{Liodakis2022,DiGesu2022}. In these models, high-energy electrons confined near the shock front would radiate in an ordered and relatively stable magnetic field (red color in Figure~\ref{fig:sketch}c), while lower-energy electrons can advect into a turbulent downstream region. Such a configuration would predict decreasing $\sigma_\text{PA}$ with increasing photon energy, opposite to the reconnection trend. However, realistic shock environments are likely to trigger turbulence both at the shock front and downstream regions, effectively erasing this distinction and yielding $\sigma_\text{PA}$ behavior similar to fully turbulent scenarios \citep{Marscher2014,Angelakis2016,ZHC2023}. Therefore, an observed rise of $\sigma_\text{PA}$ with energy uniquely supports magnetic reconnection. This physical interpretation underpins the observational comparisons presented in the following section.

The predictive power of $\sigma_\text{PA}$ extends to all types of blazars, as the synchrotron peak frequency varies across different classes (see Appendix~\ref{sec:fsrq}). For low-synchrotron-peaked (LSP) blazars, the optical band lies well above the cooling break, leading to a high $\sigma_\text{PA,O}$ in a reconnection scenario. Conversely, for and HSPs, $\sigma_\text{PA,O}$ is very low as shown in Figure~\ref{fig:polarization}. If turbulence were the primary flaring driver, $\sigma_\text{PA,O}$ would remain comparable across all blazar types; while simplified shock scenarios as the primary flaring driver would predict rising $\sigma_\text{PA,O}$ from LSPs to HSPs. This energy-dependent variability thus provides a unified benchmark for identifying the main blazar flaring mechanisms across all blazar types.

\section{Application to Multi-Wavelength Polarimetry Campaigns \label{sec:application}}

We now confront our theoretical predictions with existing multi-wavelength polarization observations of blazars. We first search for IXPE observations of HSPs that are sufficiently long and bright to collect sufficient photon statistics to characterize the polarization dynamics. This search returned five datasets: three on 1ES~1959+650 and two on Mrk~421, where polarization variability could be observed and quantified. For all observations, we use 15.125\,hour long time bins to extract polarization information, motivated by aiming to keep the measured polarization degree above or at least around the corresponding $MDP_{99}$.

For each of the five IXPE observations, we have concurrent optical R-band polarimetry data from RoboPol \citep{robopol_2019}. The cadences and number of data points vary among the optical datasets and between a given campaign's optical and X-ray data, though, as we have reported, our results are robust to these differences. We calculate the mean and standard deviation of the polarization angle and polarization degree time series and propagate the errors on each data point to get an estimate of the uncertainty associated with $\mu_\text{PD}$, $\sigma_\text{PD}$, and $\sigma_\text{PA}$. Although the statistical uncertainties remain substantial, an overall trend emerges: both $\sigma_\text{PA}$ and polarization degree variability metrics are generally higher in the IXPE band than in the optical. However, the absolute values of the observed polarization statistics do not match the single zone simulation predictions (solid lines in Figure~\ref{fig:polarization}) directly. This discrepancy is expected, as realistic blazar emission includes quiescent contribution from other parts of the blazar zone.

To facilitate a direct comparison, we supplement the simulated flaring emission with a quiescent contribution representing the broader blazar zone. This component is assumed to have a steady spectrum, a fixed polarization angle, and an energy-stratified polarization degree. This is to mimic a turbulent blazar zone that evolves on time scales much longer than the flare duration. The spectrum is assumed to be the same shape as the average turbulence spectrum in Figure~\ref{fig:xraylcsed}, and the energy-stratified polarization degree is based on the diffusion model (pink curve in Figure~\ref{fig:polarization}). Importantly, the exact spectral shape and polarization degree of this quasi-stationary component has little impact on our fitting results, as their effects can be absorbed into the relative weighting between flaring and quiescent emission. What matters is that the quiescent contribution introduces a preferred polarization angle that can suppress $\sigma_\text{PA}$, particularly at high energies.

Figure~\ref{fig:polarization} (dashed lines) shows that the reconnection scenario, when combined with such a quiescent contribution, best reproduces the observed energy dependence of $\sigma_\text{PA}$ as well as polarization degree variability metrics. In particular, reconnection intrinsically predicts low $\sigma_\text{PA}$ in the optical band, consistent with observations. At higher energies, $\sigma_\text{PA}$ is suppressed by the quiescent contribution, falling in a range consistent with IXPE observations. In contrast, a turbulence-dominated flare produces large $\sigma_\text{PA}$ at all energies, which cannot be sufficiently suppressed in the optical band without invoking unrealistically high polarization degrees in the quiescent component. This inconsistency reflects the intrinsically broad polarization angle distribution produced by turbulence.

Nonetheless, the limited statistics prevent clear quantification on whether and how much IXPE $\sigma_\text{PA}$ is higher than that in the optical band in all cases. However, the overall trend, particularly the suppression of $\sigma_\text{PA}$ in the optical band, is naturally explained by reconnection-driven flares in turbulent environment. As shown in Appendix~\ref{sec:obsdata} and Figure~\ref{fig:polratios}, the ratios between the optical and X-ray $\mu_{PD}$, $\sigma_{PD}$, and $\sigma_{PA}$ are consistent with a mixture of reconnection and turbulence. Such an interpretation is physically motivated in jets that are kink-unstable or highly magnetized, where reconnection and turbulence are expected to coexist \citep{Comisso2019,Jorstad2022,Lalakos2024}. Improved X-ray polarization statistics will be critical for fully quantifying the contributions from reconnection and turbulence.

\section{Summary and Discussion} \label{sec:summary}

In this paper, we have investigated the energy dependence of synchrotron polarization variability in blazars using PIC-integrated polarized radiation transfer simulations. Our approach self-consistently captures the co-evolution of magnetic fields and nonthermal particles in localized flaring regions. By constructing realistic pseudo-observations, we directly connect simulation predictions to multi-wavelength polarimetric data, focusing on statistical measures of polarization variability. This framework enables a robust comparison of different blazar flaring mechanisms under realistic observational conditions. Our simulations reveal several generic trends in polarization variability in a single flaring region:
\begin{enumerate}
\item Both $\mu_\text{PD}$ and $\sigma_\text{PD}$ are overall higher with higher photon energy.
\item Both $\mu_\text{PD}$ and $\sigma_\text{PD}$ start to rise sharply with photon energy around the synchrotron peak.
\item In the reconnection scenario, $\sigma_\text{PA}$ rises with photon energy till the synchrotron peak, then saturates after the peak. In contrast, the turbulence scenario predicts weak to moderate energy dependence of $\sigma_\text{PA}$, while the simplified shock scenario expects that $\sigma_\text{PA}$ generally decreases with increasing photon energy.
\item If blazar flares in all types of blazars are mainly driven by reconnection, optical $\sigma_\text{PA}$ is expected to decrease from LSPs to HSPs; if driven by turbulence, optical $\sigma_\text{PA}$ does not significantly change from LSPs to HSPs; increasing optical $\sigma_\text{PA}$ from LSPs to HSPs support the simplified shock model.
\end{enumerate}

Applying our results to existing optical and IXPE polarization campaigns for Mrk~421 and 1ES~1959+650, we find that the observed energy dependence of $\sigma_\text{PA}$ is best explained by a reconnection-driven flaring region embedded in a turbulent blazar zone. This configuration naturally arises in highly magnetized, kink-unstable jets, where reconnection and turbulence are expected to coexist \citep{Comisso2019,Jorstad2022,Agudo2025,deJonge2026}. While current IXPE data are limited by photon statistics and temporal resolution, the overall suppression of $\sigma_\text{PA}$ in the optical band favors reconnection-driven flares. Turbulence-dominated scenarios are favorable in cases where $\sigma_\text{PA}$ is comparably small at all energies. Improved X-ray polarization statistics will be essential for resolving this ambiguity.

A key advantage of our method is that it does not require strictly simultaneous observations or matched cadences across different energy bands. Instead, it relies on statistical polarization variability measured during flaring states, making it well suited to heterogeneous multi-wavelength datasets. Notably, our method implicitly assumes the data to be collected during a flaring state, which is not necessarily the case with most of the existing IXPE and optical polarization data. When no distinct flaring activity is detected, the PIC-Integrated Multi-Zone model described in \citet{deJonge2026} provides a better description of the observational data. Additionally, our simulations are restricted to 2D PIC setups. In 3D simulations, both reconnection and turbulence can show significant evolution in the third dimension \citep[e.g.,][]{Guo2021,Comisso2019,Bowyer2026}, which potentially have strong impact on the radiation and polarization signatures with different viewing angles. Therefore, future 3D simulations are needed to capture the full complexity of reconnection–turbulence interplay in relativistic jets.

The diagnostic power of energy-dependent $\sigma_\text{PA}$ opens a new avenue for probing particle acceleration and magnetic dissipation in blazar jets. Systematic optical polarimetric monitoring combined with deeper and longer IXPE observations will allow these predictions to be tested with increasing precision. More detailed analyses of spectral, flux, and polarization variability, which will be the topic of a follow-up paper, can be compared with increasingly rich multi-wavelength polarization data with higher statistics. Furthermore, trends in optical $\sigma_\text{PA,O}$ across all types of blazars offer a promising route to identifying the dominant flaring mechanism on population surveys. As multi-wavelength polarization datasets continue to grow, $\sigma_\text{PA}$ provides a physically grounded and observationally robust tool for advancing our understanding of relativistic jet physics.

\begin{acknowledgements}
We thank the anonymous referee for very insightful and constructive comments. This work is supported by IXPE GO program Cycle 1, grant number 80NSSC24K1160. HZ is supported by NASA under award number 80GSFC24M0006 and also IXPE GO program Cycle 1, grant number 80NSSC24K1173. The authors would like to thank Ioannis Liodakis for sharing the optical polarization data that we used in the paper. F. G. acknowledges the support from NSF Award 2308091. Simulations are carried out on NERSC Perlmutter cluster.
\end{acknowledgements}

\appendix 
\restartappendixnumbering

\section{Simulation Setup \label{sec:simulation}}

\begin{table}
\centering
\begin{tabular}{lcccccc} \hline\hline
Run \#  & $B_g/B_0$  & $N$  & $T_0$  & $\sigma_e$    & $C_{cool}$  & $\gamma_c$      \\ \hline
Rec1    & 0.2        & --   & 400    & $3\times10^5$ & 1600        & $6.4\times10^4$ \\ \hline
Rec2    & 0.1        & --   & 400    & $3\times10^5$ & 1600        & $6.4\times10^4$ \\ \hline
Rec3    & 0.3        & --   & 400    & $3\times10^5$ & 1600        & $6.4\times10^4$ \\ \hline
Rec4    & 0.2        & --   & 800    & $3\times10^5$ & 1600        & $6.4\times10^4$ \\ \hline
Rec5    & 0.2        & --   & 200    & $3\times10^5$ & 1600        & $6.4\times10^4$ \\ \hline
Rec6    & 0.2        & --   & 400    & $1\times10^5$ & 1600        & $6.4\times10^4$ \\ \hline
Rec7    & 0.2        & --   & 400    & $5\times10^5$ & 1600        & $6.4\times10^4$ \\ \hline
Rec8    & 0.2        & --   & 400    & $3\times10^5$ & 1200        & $6.4\times10^4$ \\ \hline
Rec9    & 0.2        & --   & 400    & $3\times10^5$ & 2000        & $6.4\times10^4$ \\ \hline
Rec10   & 0.2        & --   & 400    & $3\times10^5$ & 1600        & $1.28\times10^5$ \\ \hline
Rec11   & 0.2        & --   & 400    & $3\times10^5$ & 1600        & $3.2\times10^4$ \\ \hline \hline
Tur1    & --         & 8    & 400    & $3\times10^5$ & 1600        & $6.4\times10^4$ \\ \hline
Tur2    & --         & 4    & 400    & $3\times10^5$ & 1600        & $6.4\times10^4$ \\ \hline
Tur3    & --         & 12   & 400    & $3\times10^5$ & 1600        & $6.4\times10^4$ \\ \hline
Tur4    & --         & 8    & 800    & $3\times10^5$ & 1600        & $6.4\times10^4$ \\ \hline
Tur5    & --         & 8    & 200    & $3\times10^5$ & 1600        & $6.4\times10^4$ \\ \hline
Tur6    & --         & 8    & 400    & $1\times10^5$ & 1600        & $6.4\times10^4$ \\ \hline
Tur7    & --         & 8    & 400    & $5\times10^5$ & 1600        & $6.4\times10^4$ \\ \hline
Tur8    & --         & 8    & 400    & $3\times10^5$ & 1200        & $6.4\times10^4$ \\ \hline
Tur9    & --         & 8    & 400    & $3\times10^5$ & 2000        & $6.4\times10^4$ \\ \hline
Tur10   & --         & 8    & 400    & $3\times10^5$ & 1600        & $1.28\times10^5$ \\ \hline
Tur11   & --         & 8    & 400    & $3\times10^5$ & 1600        & $3.2\times10^4$ \\ \hline \hline
\end{tabular}
\caption{Key parameters for all reconnection (Rec) and turbulence (Tur) simulations. $B_g/B_0$ is the guide field strength for reconnection runs, which is not present for turbulence runs. $N$ is the number of the initial magnetic fluctuation modes for turbulence runs, which is not present for reconnection runs. $T_0$ is the upstream temperature, $\sigma_e$ is the electron magnetization factor, $C_{cool}$ is the cooling factor, $\gamma_c$ is the estimated cooling break location.}
\label{tab:parameter}
\end{table}

This section describes our simulation setups and assumptions and methods in our data analyses. PIC simulation setups are similar to our previous works \citep{ZHC2020,ZHC2023}, but we still include description for key parameters in the following for completeness. For magnetic reconnection, we assume a preexisting current sheet in the HSP emission region. For magnetized turbulence, we start with a uniform laminar magnetic field with a spectrum of magnetic fluctuations as the initial perturbation. Both setups are likely to present in a magnetized kink-unstable jet, which has been often seen in global magnetohydrodynamic simulations \citep{Lalakos2024,Tchekhovskoy2016}. All simulations are 2D in the $x$-$z$ plane, where $z$ is the jet bulk motion direction and the Lorentz factor is fixed at $\Gamma=10$. The viewing angle in the comoving frame is fixed at $y$ axis, so that the jet is observed at $1/\Gamma$ from the jet axis in the observer's frame and the bulk Doppler factor $\delta\equiv\Gamma=10$.

Reconnection starts from a magnetically-dominated force-free current sheet, $\mathbf{B}=B_0\tanh(y/\lambda)\mathbf{\hat{x}}+B_0\sqrt{\sech^2(y/\lambda)+B_g^2/B_0^2}\mathbf{\hat{y}}$, where $B_g$ is the strength of the guide field, which is the component perpendicular to the anti-parallel components $B_0$. The half-thickness of the current sheet is $\lambda=0.6\sqrt{\sigma_e}d_{e0}$, where $d_{e0}=c/\omega_{pe0}$ is the non-relativistic electron inertial length, $\omega_{pe0}=\sqrt{4\pi n_ee^2/m_e}$ is the non-relativistic electron plasma frequency, and $\sigma_e=B_0^2/(4\pi n_em_ec^2)$ is the cold electron magnetization parameter \citep{Sironi2014,Guo2014}. The size of the simulation box is $2L\times L$ with a resolution of $6144\times 3072$, where $L=48000 d_{e0}$.

Turbulence starts from a uniform mean magnetic field $B_0\hat{\mathbf{y}}$ and a spectrum of magnetic fluctuations $\delta\mathbf{B}$ in the $x$-$z$ plane, with $\delta B^2_\text{rms0}\equiv\left<\delta B^2\right>_{t=0}=B_0^2$ and $\delta\mathbf{B}(\mathbf{r})=\sum_{\mathbf{k}}\delta B(\mathbf{k})\hat{\mathbf{\xi}}(\mathbf{k})\exp[i(\mathbf{k}\cdot\mathbf{r}+\phi_{\mathbf{k}})]$, where $\delta B(\mathbf{k})$ is the amplitude of each wave mode, $\hat{\mathbf{\xi}}(\mathbf{k})\equiv i\mathbf{k}\times\mathbf{B}_0/|\mathbf{k}\times\mathbf{B}_0|$ is the polarization unit vector, and $\phi_{\mathbf{k}}$ is the wave phase. The wave vector $\mathbf{k}=(k_x, k_z)$, where $k_x=2\pi m/(2L)$ and $k_z=2\pi n/(2L)$ for $m,n\in\{-N,\cdots-1,1\cdots N\}$ and $N$ is the number of modes. The wave phases are assumed to be random within 0 and $2\pi$. To ensure that $\delta\mathbf{B}$ is real, we assume $\delta B(-\mathbf{k})=\delta B(\mathbf{k})$ and $\phi_{-\mathbf{k}}=-\phi_{\mathbf{k}}$. If each wave mode carries the same power \citep[equal amplitude per mode, similar to][]{Comisso2019,ZHC2023,Singh2025}, $\delta B(\mathbf{k})=\delta B_\text{rms0}/2N$. The simulation box size is $2L\times 2L$ with a resolution of $6144\times 6144$, where $L=48000 d_{e0}$ as well.

Other than $B_g$ and $N$ that are specific for reconnection and turbulence, respectively, all other parameters that we survey are exactly the same for the two mechanisms to facilitate direct comparisons between the two scenarios. We assume an electron-proton plasma with realistic mass ratio $m_i/m_e=1836$. The initial particle distributions are Maxwell–J\"uttner distributions with a uniform density $n_0$ and temperature $T_0$ for both electrons and protons. For the fiducial parameter set Run 1 for reconnection and turbulence, $T_0=400 m_e c^2$. The upstream thermal electron inertial length is then $d_e=\sqrt{1+3T_0/(2m_ec^2)}d_{e0}\sim 24.5d_{e0}$, which is resolved by our simulations. We choose $\sigma_e=3\times10^5$ that allows the electron spectrum to reach the typical synchrotron spectral peak within a reasonably short period of simulation time. The total magnetization in the simulation box is approximately $\sigma\sim \sigma_e m_e/m_i$. Each PIC cell has $100$ electron-ion pairs. We implement a radiation reaction force for the synchrotron cooling effect, which can be considered as a continuous frictional force for relativistic electrons \citep[non-relativistic terms are ignored; see][]{Cerutti2012,Cerutti2013},
\begin{align}
  \mathbf{g}
  = -\frac{2}{3}r_e^2\gamma\left[\left(\mathbf{E}+
  \frac{\mathbf{u}\times\mathbf{B}}{\gamma}\right)^2 -
  \left(\frac{\mathbf{u}\cdot\mathbf{E}}{\gamma}\right)^2\right]\mathbf{u},
\end{align}
where $\mathbf{u}=\gamma \mathbf{v}/c$ is the four-velocity and $r_e=e^2/m_ec^2$ is the classical radius of the electron. Compton scattering cooling is ignored in this paper, but as shown in \citet{ZHC2022}, its effects on radiation and polarization signatures in the synchrotron component are minimal. Given the fact that the typical blazar cooling parameters have trivial effects on PIC scales, we compensate the cooling force by a factor $C_{cool}$ so that the cooling break of the particle spectrum occurs at $\sim \gamma_c$. Table~\ref{tab:parameter} lists all parameters that we survey in this paper.

 We further limit our simulation time to be three light crossing time scales ($3\tau_{lc}$), which, if extrapolated to the typical blazar zone size of $10^{16-17}\, \rm{cm}$, correspond to a flaring episode of a few days to a couple of weeks \citep{Boettcher2013}. This ensures that the radiation and polarization signatures are dominated by the local particle acceleration. Otherwise the large-scale jet dynamics will inevitably affect the flux and polarization variations on longer time scales. 

The magnetic field and particle evolution obtained from \texttt{VPIC} \citep{Bowers2008} are post-processed with the \texttt{3DPol} code \citep{ZHC2018}. The initial magnetic field strength is normalized to $0.1~\rm{G}$, which is a typical value for the leptonic scenario \citep[e.g.,][]{Boettcher2013}. To obtain adequate statistics of particle spectra for the radiative transfer, we collect all electrons in $8\times 8$ PIC cells and obtain their spectrum, and calculate the average magnetic field within these PIC cells. This is then considered as one cell in the radiation transfer. In this way, we include all particles in PIC, but any disorder of the magnetic field components on scales smaller than the radiative transfer cell are down-sampled. We bin the particle kinetic energy $(\gamma_e-1)m_ec^2$ into 100 steps between $10^{-4}m_ec^2$ and $10^6m_ec^2$. PIC simulations last $3\tau_{lc}$, where $\tau_{lc}$ is the light crossing time in the $x$-axis, and we output the above information every $\sim 0.005\tau_{lc}$, in total 600 time steps to obtain adequate temporal resolution. \texttt{3DPol} then calculates the Stokes parameters at every time step in each radiative transfer cell, and ray-traces to the plane of the sky.

\section{Data Analysis \label{sec:analysis}}

\begin{figure*}
\centering
\includegraphics[width=0.48\linewidth]{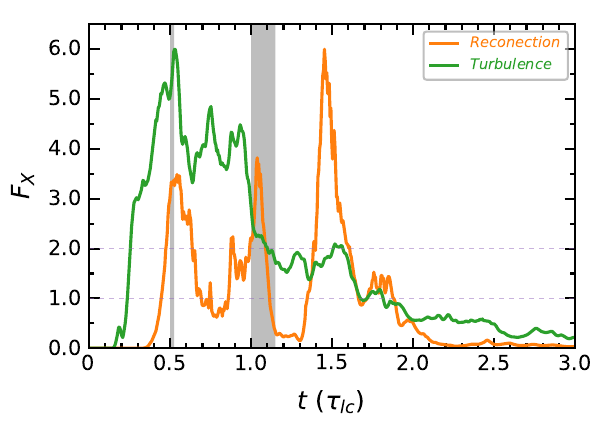}
\includegraphics[width=0.48\linewidth]{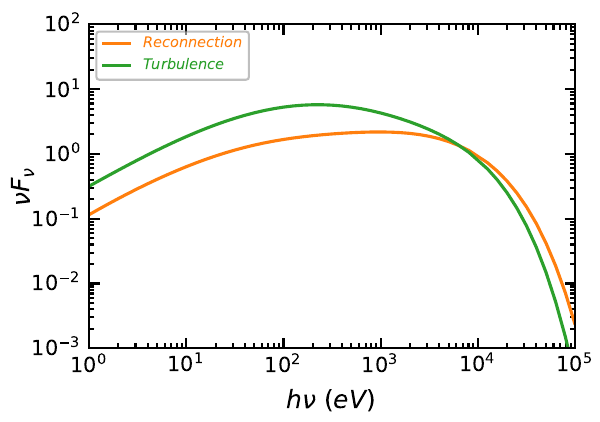}
\caption{Left Panel: 2-8 keV light curves for Run 1 of magnetic reconnection (orange) and magnetized turbulence (green). Light curves are normalized to a maximal flux of 6 and have 600 time steps. The gray horizontal dashed lines represent the two threshold selections, while the vertical shaded regions illustrate the size of the time window selections. Right Panel: The average spectra of the Run 1 of magnetic reconnection (orange) and magnetized turbulence (green) simulations.}
\label{fig:xraylcsed}
\end{figure*}

This section describes our data analysis assumptions and methods. We normalize the maximal flux of 2-8 keV band in every run to 6 so that the maximal flaring amplitudes in the IXPE band is the same to facilitate direct comparison between runs and observations. We consider different source and observational conditions that can affect the observable signatures in the data analyses.

Figure~\ref{fig:xraylcsed} left panel shows the 2-8 keV light curve for Rec1 and Tur1, and the right panel shows their average synchrotron spectral components. It is well known that blazars can have quiescent and flaring states, and the quiescent and part of the flaring emission may come from more than one regions of the jet. Our coupled simulations, however, only represent one active flaring region. Therefore, when the flux from this region is low, the total blazar emission is likely dominated by other regions in the jet. Additionally, due to limited sensitivity, telescopes may not be able to capture low flux from the flaring region even if it dominates over emission from all other parts of the jet. Therefore, we choose a threshold (1 or 2) in the 2-8 keV light curves that are normalized to a maximal flux of 6, below which all simulation results are ignored. In this way, we only consider detectable emission from one dominating flaring region that we simulate. Moreover, observations have gaps and various cadence due to sensitivity, scheduling, and many other factors. But our simulated data is continuous and can have arbitrarily high temporal resolution in principle. We choose a cadence in the 2-8 keV light curve: if 5 or 30 time steps of the simulated 2-8 keV light curve are all above the chosen threshold, we collect one data point. For example, the 2-8 keV light curve of Rec1 has four epochs above the threshold, each lasts 30, 57, 3, and 26 code time steps. Then for a cadence of 5 time steps, we get 6, 11, 0, and 5 data points for this run. Each data point contains the multi-wavelength Stokes parameters averaged over the 5 time steps in the time window, thus depolarization due to variable polarization angle within the unresolved 5 time steps is included, mimicking a real observation. Although our choices of thresholds and cadences and normalization of maximal 2-8 keV flux to 6 do not match real observations, they are intended to examine the robustness of our results with different flare amplitudes, observational sensitivity and cadences, and asynchronous data.

\section{Observational Data \label{sec:obsdata}}

\begin{figure*}
\centering
\includegraphics[width=0.32\linewidth]{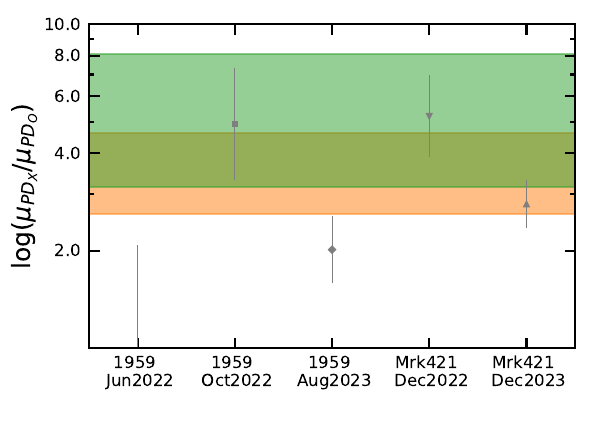}
\includegraphics[width=0.32\linewidth]{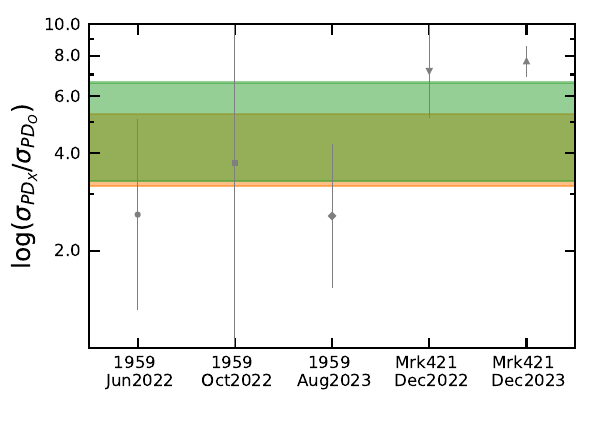}
\includegraphics[width=0.32\linewidth]{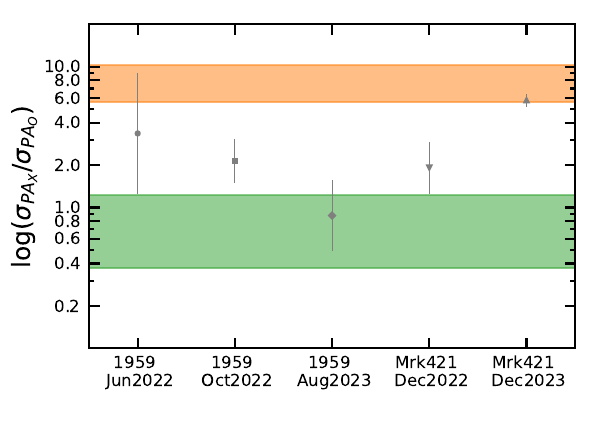}
\caption{From left to right are ratios of $\mu_{PD}$, $\sigma_{PD}$, and $\sigma_{PA}$, respectively, between the IXPE and optical bands. Gray data points are the five observations in Table~\ref{tab:ixpedata}, and the orange and green shades represent the reconnection and turbulence model predictions, respectively.}
\label{fig:polratios}
\end{figure*}

Table~\ref{tab:ixpedata} shows the X-ray data plotted in Figure~\ref{fig:polarization} for five IXPE observations of the two High Synchrotron Peaked blazars 1ES~1959+650 and Mrk~421. These are IXPE observation IDs 01006001, 02004801, 02250801, 02004401, and 02008199. Error bars represent the errors on individual data points propagated through the relevant formula. The average polarization degree $\mu_{PD}$ is not the proper average from determining the polarization degree from the net Stokes parameters for the whole observation, but instead the statistical mean of the binned polarization degrees over the course of the observation. This is done this way so that the calculation of the average polarization degree (in addition to the standard deviations of the polarization degree and angle) is consistent with the method for calculating the polarization degree in the simulations. Figure~\ref{fig:polratios} shows the ratios of $\mu_{PD}$, $\sigma_{PD}$ and $\sigma_{PA}$ between IXPE and optical bands. The ranges of reconnection and turbulence models are obtained by assuming the flux ratio of the quiescent region over the flaring region spans over 0.1 to 3, and the quiescent polarization degree in the IXPE band ranges from 0.15 to 0.4. It further shows that $\mu_{PD}$ and $\sigma_{PD}$ cannot clearly disentangle reconnection and turbulence scenarios, as the two models predict overlapping ratios, and existing observations are generally consistent with both scenarios. Only $\sigma_{PA}$ shows clear difference between reconnection and turbulence, and the Mrk~421 observation on Dec 2023 is consistent with reconnection-driven flare, while 1ES~1959+650 observation on Aug 2023 is consistent with turbulence-driven flares.

\begin{table}
\centering
\begin{tabular}{cccc}
\hline \hline
OBSID    & $\mu_{PD} (\%)$ & $\sigma_{PD} (\%)$ & $\sigma_{PA} (^{\circ})$  \\ 
\hline
01006001 & $32.0 \pm 3.0$ & $2.59 \pm 1.74$  & $28.3 \pm 28.1$          \\ 
\hline
02004801 & $12.5 \pm 5.0$ & $1.53 \pm 2.74$  & $17.2\pm 5.95$          \\ 
\hline
02250801 & $10.4 \pm 2.5$ & $1.91 \pm 0.98$  & $6.06\pm 3.53$          \\ 
\hline
02004401 & $15.0 \pm 4.3$ & $5.87 \pm 1.95$  & $17.1\pm 7.26$          \\ 
\hline
02008199 & $12.5 \pm 2.1$ & $6.50 \pm 0.59$  & $23.1\pm 1.68$ 
\\
\hline \hline
\end{tabular}
\caption{\textit{IXPE} HSP Polarization Data}
\label{tab:ixpedata}
\end{table}

\section{Robustness Tests \label{sec:robustness}}

\begin{figure*}
\centering
\includegraphics[width=0.32\linewidth]{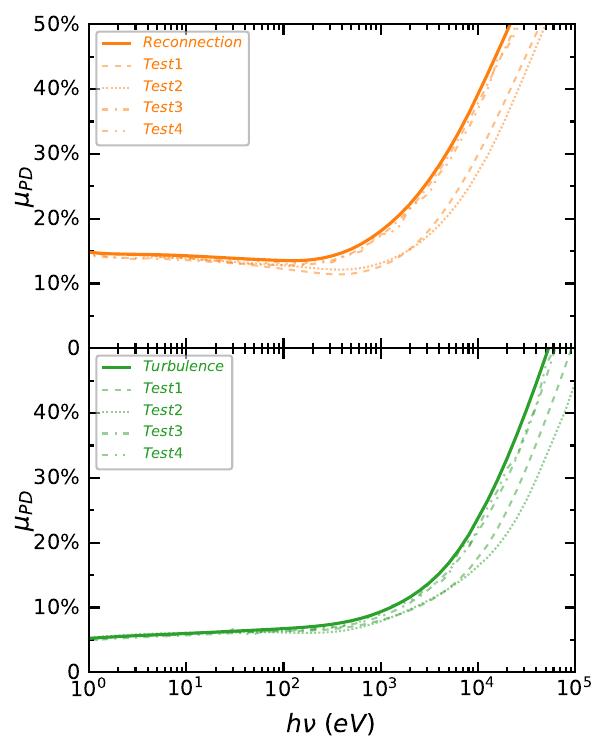}
\includegraphics[width=0.32\linewidth]{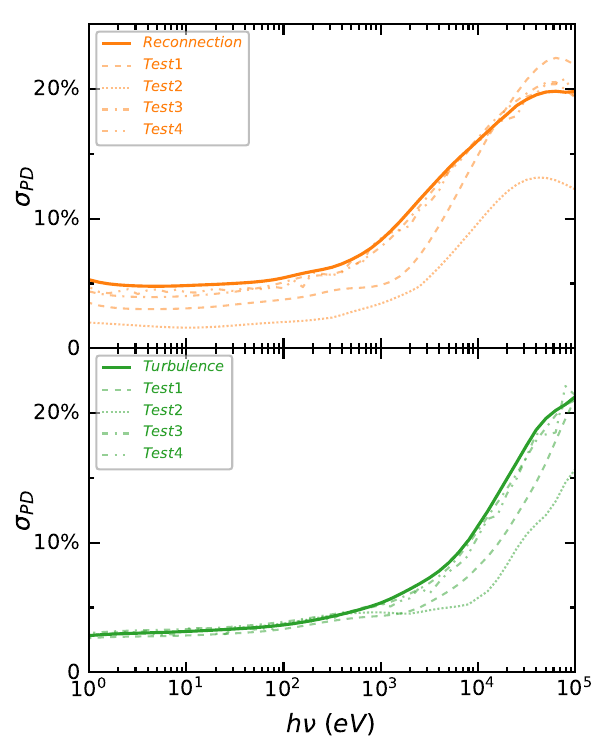}
\includegraphics[width=0.32\linewidth]{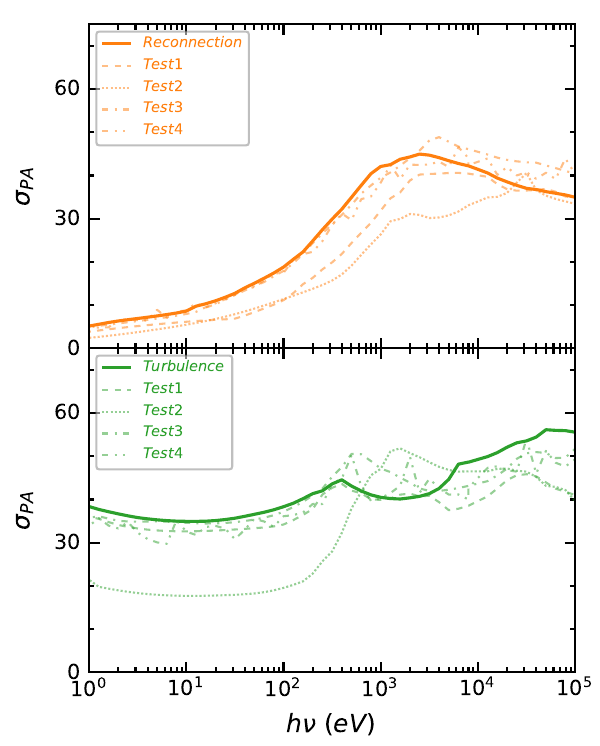}
\caption{Energy stratification of polarization for reconnection (orange) and turbulence (green) with different thresholds, cadences, and/or asynchronous data points. From left to right are $\mu_\text{PD}$, $\sigma_\text{PD}$, and $\sigma_\text{PA}$. In each plot the upper panel is reconnection and the lower is turbulence. Solid lines are the same as in Figure~\ref{fig:polarization}. Dashed lines (Test 1) use threshold of 2 and cadence of 5. Dotted lines (Test 2) use threshold of 2 and cadence of 30. Dash-dotted lines (Test 3) and dash-dot-dotted lines (Test 4) use random choices of 30 and 10 asynchronous data points per run, respectively, with arbitrary thresholds and cadences. The blue and pink curves in the left panel represent the semi-analytical models described in \citet{Angelakis2016} and \citet{ZHC2024}, respectively.}
\label{fig:robustness}
\end{figure*}

Figure~\ref{fig:robustness} presents our robustness test. Solid lines are the fiducial threshold 1 cadence 5 case; each run (flaring episode) on average has $\gtrsim 60$ data points for reconnection and $\gtrsim 100$ data points for turbulence. Test 1 is the threshold 2 cadence 5 case; each run has on average $\sim 20$ data points for reconnection and $\gtrsim 50$ data points for turbulence. Test 2 is the threshold 2 cadence 30 case; each run has $\sim 5$ data points for reconnection and $10$-$\gtrsim 20$ data points for turbulence. Tests 3 and 4 are obtained in the following way. First, we collect all data points for different thresholds and cadences, i.e., threshold 1 cadence 5, threshold 1 cadence 30, threshold 2 cadence 5, and threshold 2 cadence 30. Then we randomly draw 30 and 10 data points per run in the optical band for Tests 3 and 4, respectively. Repeat the previous step with a different set of random selections in the IXPE band. The data points thus can be drawn from any combination of thresholds and cadences, and the optical and X-ray data points are asynchronous. We calculate $\mu_\text{PD}$, $\sigma_\text{PD}$, and $\sigma_\text{PA}$ based on the selected data points.

\section{Flat Spectrum Radio Quasar \label{sec:fsrq}}

\begin{figure*}
\centering
\includegraphics[width=0.48\linewidth]{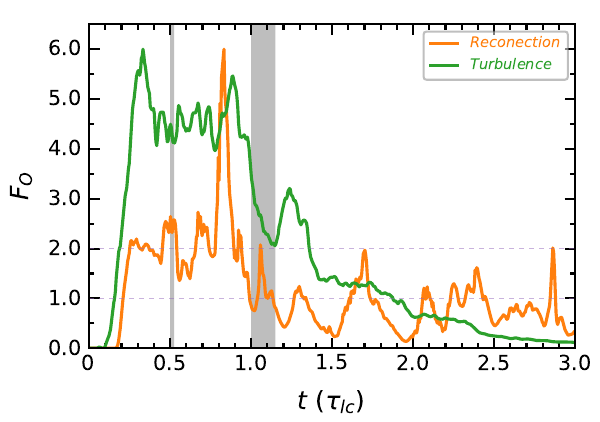}
\includegraphics[width=0.48\linewidth]{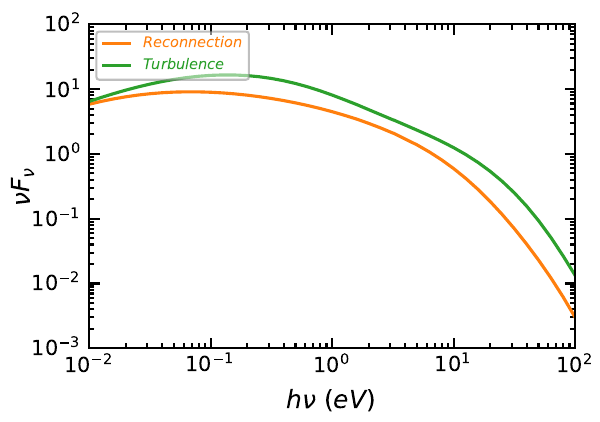}
\caption{Left Panel: optical light curves for Run 1 of magnetic reconnection (orange) and magnetized turbulence (green). Light curves are normalized to a maximal flux of 6 and have 600 time steps. Thresholds and cadences are the same as in Figure~\ref{fig:fsrqlcsed}. Right Panel: The average spectra of the Run 1 of magnetic reconnection (orange) and magnetized turbulence (green) simulations.}
\label{fig:fsrqlcsed}
\end{figure*}

\begin{figure*}
\centering
\includegraphics[width=0.32\linewidth]{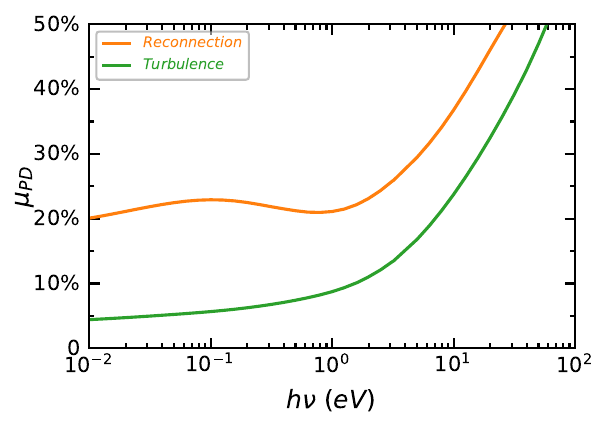}
\includegraphics[width=0.32\linewidth]{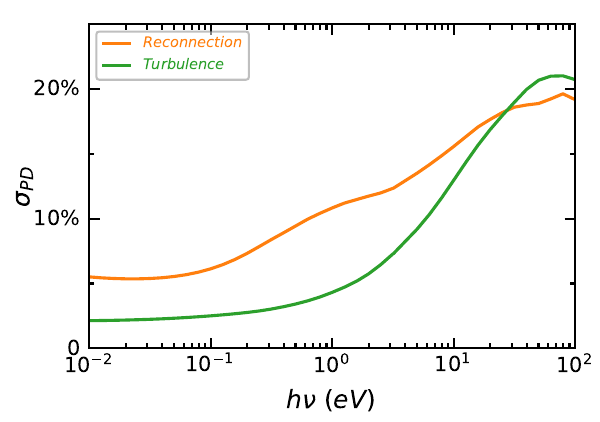}
\includegraphics[width=0.32\linewidth]{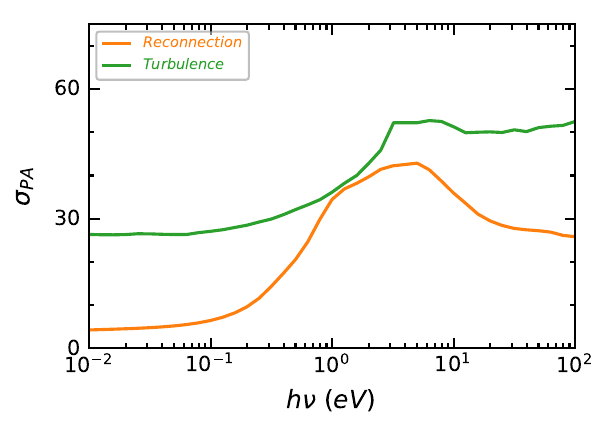}
\caption{Same as Figure \ref{fig:polarization} but for FSRQs.}
\label{fig:fsrq}
\end{figure*}

Figure~\ref{fig:fsrqlcsed} shows an example of optical light curves and average spectra for FSRQ simulations. Figure~\ref{fig:fsrq} shows the energy stratified polarization signatures. Parameters are summarized in Table~\ref{tab:fsrqparameter}. It is clear that all trends described for the HSP blazars apply similarly for the FSRQs, except that the spectral peak now moves to the near infrared bands.

\begin{table}
\centering
\begin{tabular}{lcccccc} \hline\hline
Run \#  & $B_g/B_0$  & $N$  & $T_0$  & $\sigma_e$    & $C_{cool}$  & $\gamma_c$      \\ \hline
Rec1    & 0.2        & --   & 40     & $2\times10^4$ & 200         & $2\times10^3$   \\ \hline
Rec2    & 0.2        & --   & 40     & $1\times10^4$ & 200         & $2\times10^3$   \\ \hline
Rec3    & 0.2        & --   & 40     & $4\times10^4$ & 200         & $2\times10^3$   \\ \hline
Rec4    & 0.1        & --   & 40     & $2\times10^4$ & 200         & $2\times10^3$   \\ \hline
Rec5    & 0.3        & --   & 40     & $2\times10^4$ & 200         & $2\times10^3$   \\ \hline
Rec6    & 0.2        & --   & 40     & $2\times10^4$ & 100         & $2\times10^3$   \\ \hline
Rec7    & 0.2        & --   & 40     & $2\times10^4$ & 400         & $2\times10^3$   \\ \hline
Rec8    & 0.2        & --   & 40     & $2\times10^4$ & 200         & $1\times10^3$   \\ \hline
Rec9    & 0.2        & --   & 40     & $2\times10^4$ & 200         & $4\times10^3$   \\ \hline \hline
Tur1    & --         & 8    & 40     & $2\times10^4$ & 200         & $2\times10^3$   \\ \hline
Tur2    & --         & 8    & 40     & $1\times10^4$ & 200         & $2\times10^3$   \\ \hline
Tur3    & --         & 8    & 40     & $4\times10^4$ & 200         & $2\times10^3$   \\ \hline
Tur4    & --         & 4    & 40     & $2\times10^4$ & 200         & $2\times10^3$   \\ \hline
Tur5    & --         & 12   & 40     & $2\times10^4$ & 200         & $2\times10^3$   \\ \hline
Tur6    & --         & 8    & 40     & $2\times10^4$ & 100         & $2\times10^3$   \\ \hline
Tur7    & --         & 8    & 40     & $2\times10^4$ & 400         & $2\times10^3$   \\ \hline
Tur8    & --         & 8    & 40     & $2\times10^4$ & 200         & $1\times10^3$   \\ \hline
Tur9    & --         & 8    & 40     & $2\times10^4$ & 200         & $4\times10^3$   \\ \hline \hline
\end{tabular}
\caption{Same as Table~\ref{tab:parameter} but for all reconnection (Rec) and turbulence (Tur) simulations for FSRQs.}
\label{tab:fsrqparameter}
\end{table}

\bibliography{Blazar}{}
\bibliographystyle{aasjournalv7}



\end{document}